\documentclass[prl,twocolumn,showpacs,preprintnumbers,amsmath,amssymb]{revtex4}

\usepackage{graphicx}
\usepackage{dcolumn}
\usepackage{bm}
\usepackage{units}
\usepackage{color}
\usepackage{amsmath}

\begin{document}


\author{Harald Wunderlich}
 \email{harald.wunderlich@uni-ulm.de}
\affiliation{Institut f{\"u}r Theoretische Physik, Universit{\"a}t Ulm, 
Albert-Einstein Allee 11, D-89069 Ulm, Germany}
\author{Shashank Virmani}
\affiliation{SUPA, Physics Department, University of Strathclyde, United Kingdom}
\author{Martin B. Plenio}
\affiliation{Institut f{\"u}r Theoretische Physik, Universit{\"a}t Ulm, 
Albert-Einstein Allee 11, D-89069 Ulm, Germany}

\date{\today}
\title{Highly-efficient estimation of entanglement measures for large experimentally created graph states via simple measurements}

\begin{abstract}
Quantifying experimentally created entanglement could in principle be accomplished by measuring the entire density matrix and calculating an entanglement measure of choice thereafter. Due to the tensor-structure of the Hilbert space, this approach becomes infeasible even for medium-size systems.
Here we present methods to quantify the entanglement of arbitrarily large two-colorable graph states from simple measurements. The measurement data considered here is merely given by stabilizer measurements, thus leading to an exponential reduction in the number of measurements required. We provide analytical results for the robustness of entanglement and the relative entropy of entanglement.
\end{abstract}


\maketitle

\paragraph{Introduction}
Detecting \cite{GT09} and quantifying entanglement \cite{PV07} is one of the major tasks in quantum information science. Experimentally created entanglement can in principle be quantified by determining the quantum state via full tomography, and calculating an entanglement measure of choice for this state. Apart from exceptions like the negativity, entanglement measures usually involve optimization problems, which makes them hard to calculate. Another issue is the tensor-structure of the Hilbert space, which implies that the number of measurement settings grows exponentially with the number of constituents involved in the system. Despite the recent developments in efficient tomography \cite{CP10}, the determination of the full quantum state appears to involve an unnecessary overhead given that only a single number, the value of the entanglement measure, is required. For this reason, more sophisticated methods for the direct quantification of entanglement in many-body systems are required.

Here we present direct and experimentally efficient methods to quantify entanglement of quantum many-body systems. We put the emphasis on two-colorable graph states, which represent a vast resource for applications in quantum information science. They encompass Greenberger-Horne-Zeilinger (GHZ) states \cite{Greenberger89}, Calderbank-Shor-Steane (CSS) error correction codeword states, and cluster states \cite{Briegel01}. Due to the importance of graph states, a considerable experimental effort has been made to realize them using photons \cite{Walther05, Kiesel05, Lu07, Chen07, Vallone08}, cold atoms \cite{Mandel03}, and proposals for trapped ions are pursued  \cite{WWSSK09,Stock09,Ivanov08}.

We will show that the entanglement - according to a variety of entanglement measures \cite{PV07} - of such two-colorable graph states can be estimated efficiently via measurements of the stabilizer operators only, thus reducing the experimental effort in measuring the state exponentially. Furthermore our method of entanglement estimation is purely analytic, thus avoiding computationally costly post-processing of measurement data.

\paragraph{Entanglement estimation}
Graph states of $n$ qubits correspond to a graph $G$ of $n$ vertices, with $n$ binary indices $(k_1,\dots,k_n)$. We denote the Pauli matrices at the $i$-th qubit by $X_i,Y_i,Z_i$. One can show that the $2^n$ graph states $|G_{k_1,\dots,k_n}\rangle$ are the simultaneous eigenstates of the $n$ mutually commuting operators:
$K_i := X_i \bigotimes_{Ngb(i)} Z_j$, $i=1,\dots,n$,
where $Ngb$ denotes the set of all neighbours of qubit $i$ defined by the graph.
Graph states satisfy the following eigenvalue equation:
$K_i |G_{k_1,\dots,k_n}\rangle = (-1)^{k_i}|G_{k_1,\dots,k_n}\rangle$.
The $n$ operators $K_i$ generate an abelian group $\mathcal{S}$, called the {\it stabilizer}. An experimentally created graph state could in principle be verified by measuring the $2^n$ elements of the stabilizer. As mentioned in the introduction, full-state tomography is not an option for determining the properties of a quantum many-body system due to an exponentially fast growing measurement effort. We will see that merely the measurement results of the generators of the stabilizer suffice to attain highly useful bounds on entanglement measures.

Let us suppose the goal of an experiment is the creation of a two-colorable graph state, and the generators of the stabilizer are measured with outcomes $a_i = tr(\rho K_i)$, $i=1,\dots,n$. As convention for the coloring we use $|A|$ Amber and $|B|$ Blue qubits, taking $|A| \ge |B|$. A generator $K_i$ is said to be Amber (Blue) if $i$ corresponds to an Amber (Blue) qubit.

Given this tomographically incomplete data, one is now interested in finding the minimal entanglement (according to a certain entanglement measure) compatible with the measurement data. Mathematically, this is a formulated as the semidefinite program  \cite{AP06, WP0902}:
\begin{equation}
\label{eq:Emin}
E_{min} = min_{\rho} \{E(\rho) : tr(\rho K_i)=a_i , \rho \ge 0\},
\end{equation} 
where $E(\rho)$ is the entanglement quantifier of choice. We will consider the following entanglement measures: the {\it relative entropy of entanglement} is defined as \cite{VP98}
\begin{equation}
E_R(\rho) = min_{\sigma \in SEP} tr[\rho (log_2 \rho - log_2 \sigma),
\end{equation}
where $SEP$ denotes the set of fully separable states. The {\it global robustness of entanglement} is given by the minimum amount of an unnormalized state that has to be mixed in to the given state to wash out all entanglement \cite{RobustnessEnt}:
\begin{equation}
R(\rho) = min_{\sigma} \{ tr(\sigma) : \rho+\sigma \in SEP\}.
\end{equation}


Let us return to the estimation of entanglement measures from stabilizer measurements. The crucial point is that the minimization (\ref{eq:Emin}) does not need to be carried out over all states $\rho$. Instead, it suffices to minimize the entanglement quantifier over stabilizer diagonal states only. This can be seen in the following way: since the stabilizer operators are mutually commuting, the measurement outcomes $tr(\rho K_i)$ are invariant under any rotation of the density matrix of the form $\rho \rightarrow K_j \rho K_j$, $K_j \in \mathcal{S}$. Due to convexity of the entanglement quantifier, it is legitimate to apply a local symmetrization procedure ,colloquially referred to as "twirling", to the state. This is performed by averaging over all stabilizer rotations $\rho \rightarrow \frac{1}{2^n} \sum_{j=1}^{2^n} K_j \rho K_j$. In so doing, the optimization is restricted to stabilizer diagonal states of the form
\begin{equation}
\rho = \frac{1}{2^n} \sum_{i_1,\dots,i_n=0}^1 c_{i_1,\dots,i_n} K_1^{i_1}\dots K_n^{i_n},
\end{equation}
where some coefficients are determined by the measurement outcomes, while the rest are variables. Since the stabilizer operators are mutually commuting, and their spectrum  is given by $\{-1,+1\}$, it is straightforward to compute the eigenvalues of $\rho$ as
$ \lambda_{\vec{j}} = \frac{1}{2^n} \sum_{\vec{i}} (-1)^{\vec{i}\cdot \vec{j}} c_{\vec{i}} $. 
Note that a stabilizer diagonal state corresponds to a mixture of graph states generated by these stabilizer operators. This can easily be seen in the following way:
\begin{align}
\rho & = \sum_{\vec{k}} \lambda_{\vec{k}} |G_{\vec{k}}\rangle \langle G_{\vec{k}}| 
= \sum_{\vec{k}} \lambda_{\vec{k}} \frac{1}{2^n} \sum_{\vec{i}} (-1)^{\vec{i}\cdot{k}} K_1^{i_1} \dots K_n^{i_n} \\
& =   \sum_{\vec{i}} c_{\vec{i}} K_1^{i_1} \dots K_n^{i_n}, \text{with}~c_{\vec{i}} =  \frac{1}{2^n} \sum_{\vec{k}} (-1)^{\vec{i}\cdot{\vec{k}}} \lambda_{\vec{k}}.
\end{align}

\paragraph{Estimating the robustness of entanglement} In order to estimate the global robustness of entanglement, we begin by bounding it from below in the following way: for any mixed state $\rho = \sum_{k} \lambda_k \rho_k$ and any index $m$ it holds that \cite{MMV07}:
\begin{equation}
\label{eq:Rmix}
R(\rho) \ge \lambda_m (1+R(\rho_m))-1.
\end{equation}
Since we minimize the robustness over twirled states of the form $\rho = \sum_{\vec{k}} \lambda_{\vec{k}} |G_{\vec{k}}\rangle \langle G_{\vec{k}}|$, a lower bound on the robustness of entanglement consistent with the stabilizer measurements is provided by the minimum fidelity $F$ that may be inferred from such measurements.
The minimization of the least fidelity compatible with stabilizer measurements reads: $F  = min_{\rho} [tr(\rho |G_{(k_1,\dots,k_n)}\rangle  \langle G_{(k_1,\dots,k_n)}| ) : tr(\rho K_i)=a_i, \rho \ge 0 ]$. By Lagrange duality, one finds the dual problem: $F  = max_{\mu_i}[\sum_i \mu_i a_i : 
|G_{(k_1,\dots,k_n)}\rangle  \langle G_{(k_1,\dots,k_n)}| - \sum_i K_i \ge 0]$. Solutions to primal and dual can be constructed for an arbitrary number of qubits, attaining the following analytical optimal solution \cite{WP0902}: $F  = max[0, \frac{1}{2}(\sum_i a_i - n + 2)]$. Combining this with Eq. (\ref{eq:Rmix}) provides us with the following lower bound on the global robustness of entanglement that can be achieved from stabilizer measurements:
\begin{equation}
R_{min}(\rho) \ge max\{0, 2^{|B|} max[0, \frac{1}{2}(\sum_i a_i - n + 2)]-1 \}.
\end{equation}

\paragraph{Estimating the relative entropy} In a similar fashion, we now calculate the lower bound on the relative entropy on entanglement in the case of stabilizer measurements. First, note that once more the optimization may be restricted to stabilizer diagonal states resp. mixtures of graph states. Then, the lower bound on the relative entropy is given by
\begin{equation}
{E_R}_{min} (\rho) \ge max\{0, |B|-max[S(\rho): tr(\rho K_i)=a_i,\rho \ge 0]\}.
\end{equation}
We can prove this in the following way: first, note that the two-coloring divides the system in the two partitions $A$ and $B$. Now one uses the fact that the relative entropy is lower bounded by the difference between the entropy of system $A$ resp. $B$ and the entropy of the total system \cite{PVP00}:
\begin{equation}
E_R (\rho_{AB}) \ge max(S(\rho_A),S(\rho_B))-S(\rho_{AB}).
\end{equation}
In our case, we consider only mixtures of two-colorable graph states, so that tracing out system $A$ results in a maximally mixed state with entropy $|B|$.
Hence, the minimization of the relative entropy involves an entropy maximization. This can be achieved as outlined in \cite{WP0907}: measuring the generators of the stabilizer group gives rise to probability distribution $p_k^{(\pm)} = \frac{1\pm a_k}{2}$ for the projections upon the stabilizer eigenspaces. Furthermore, we denote the probability distribution of the joint state of the system by $\lambda_{i_1 \dots i_n}$. A crucial feature of the entropy is subadditivity:
$S(\lambda_{i_1 \dots i_n}) \le \sum_{k=1, s=\pm}^n H(p_k^{(\pm)})$, where $H$ denotes the classical entropy function. 
A little thought shows that the above inequality holds with equality for the probability distribution given by
$\lambda_{i_1 \dots i_n} = \prod_{k=1}^n \frac{1+(-1)^{i_k} a_k}{2}$, 
thus giving the exact maximal entropy $S_{max} = -\sum_{i_1 \dots i_n=0}^1 \lambda_{i_1 \dots i_n} log \lambda_{i_1 \dots i_n}$. To conclude, the lower bound on the relative entropy of entanglement that can be inferred from stabilizer measurements is computed as
\begin{equation}
{E_R}_{min} = max\{0, |B|+\sum_{i_1 \dots i_n=0}^1 \lambda_{i_1 \dots i_n} log \lambda_{i_1 \dots i_n} \}.
\end{equation}
%
%
\paragraph{Upper bounds} In the previous paragraphs we have derived lower bounds on the
minimal entanglement that is consistent with the statistics obtained from measuring
individual stabilizer operators. It is also of use to derive upper bounds. A simple approach to doing this is available using the results
of \cite{MMV07}. The total Hilbert space can be divided up into subspaces, where each subspace is
labelled by a deterministic outcome for all the Amber stabilizers. For any pure two colourable graph state the entanglement
$E_R$ is given by $|B|$, and the robustness of entanglement is given by $2^{|B|}-1$. In \cite{MMV07}
it was shown that for a mixed (twirled) state supported entirely in such a subspace, $E_R$ is given by
$E_R(\rho)=|B|-S(\rho),$
whereas
$R(\rho)=2^{|B|}\max \lambda_{\vec{k}}-1$.
Let us use the symbol $a$ to denote a possible set of outcomes for the Amber measurements,
and let $b$ denote a possible set of outcomes for the Blue measurements.
Hence any state that is diagonal in the graph state basis can be described as a
probability distribution $p(a,b)=p(a)p(b|a)$ corresponding to the probabilities for getting
the various possible stabilizer outcomes. We can partition such a state into
a mixture of states that are individually supported on each of the Amber subspaces, such that $\rho = \sum p(a) \rho_{a}$, where $a$ is a bit string corresponding to the positive/negative stabilizer subspaces of the Amber qubits. 
By concavity of the entropy function we find that: $E_R(\rho) \le |B|-\sum_a p(a) S(\rho_{a})$. But $S(\rho_{a})$ is given by a classical entropy $H( p(b|a))$  where $p(b|a)$ is the conditional probability distribution for getting outcomes $b$
upon finding $a$. Thus we obtain:
\begin{equation}
E_R(\rho) \leq |B|- \sum_a p(a) H(p(b|a)).
\end{equation}
Similarly, since the $A$ subspace entanglement is given by $R(\rho_{a}) = 2^{|B|} (\lambda_{max}(\rho_a)-1)$, we find
\begin{equation}
R(\rho)+1 \leq 2^{|B|}\sum_a p(a) max_b p(b|a).
\end{equation}

Hence to get upper bounds to the minimal entanglement consistent
with the measurement outcomes, we need to pick the $p(a,b)$ consistent
with the marginal distributions that minimises these expressions. The relative entropy can now be estimated by noticing that the conditional entropy is upper bounded by $H(p(b))$ and choosing a product distribution $p(b)=p(b_1)...p(b_{|B|})$, for which it is well known that it maximises $H(p(b))$. Thus, we obtain
\begin{equation}
E_{R_{min}}(\rho) \leq  |B| - H(p(b)).
\end{equation}
Let $b^*$ be the maximiser of $p(b)$. As $b^*$ is a specific value of $b$, it holds that
$p(a)\max_b p(b|a) \geq p(a)p(b) = p(a,b^*)$. It follows that $\sum_a p(a)\max_b p(b|a) \geq p(b^*)$, a lower bound that is tight as it is attained by setting $p(a,b)=p(a)p(b)$.
Since we do not have complete information about $p(b^*)$, we need to minimise it subject to the individual stabilizer statistics $p(b_1)$,...,$p(b_{|B|})$. This problem is equivalent to minimising the $l_{\infty}$ norm of a joint probability distribution $p(b)=p(b_1,...,b_{|B|})$ constrained to fixed marginals $p(b_1)$,...,$p(b_{|B|})$, which is equivalent to the fidelity minimisation (only considering the Blue stabilizers). Hence we obtain:
\begin{equation}
R_{min}(\rho)+1 \leq 2^{|B|} max \left( 0, \frac{1}{2} \left(\sum_{i \in B} a_i - |B|+2\right)\right).
\end{equation}

\begin{figure}
\includegraphics[width=0.5\textwidth]{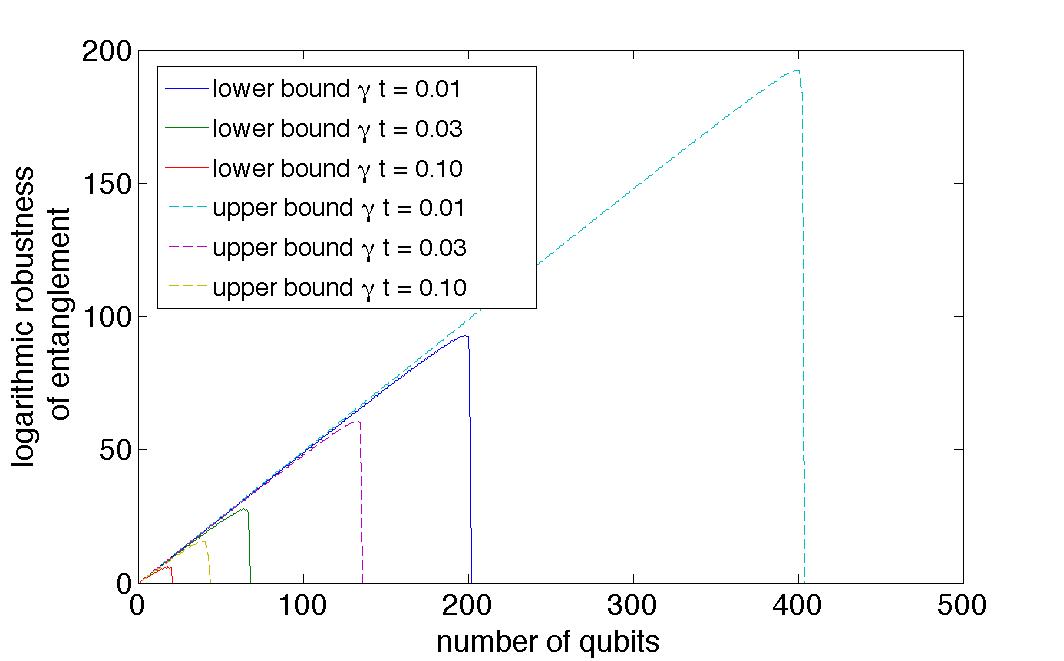}
\caption{Upper and lower bounds on the estimate of the logarithmic global robustness of entanglement for linear graph states subject to local dephasing. A non-zero estimate of the entanglement is possible as long as a non-zero fidelity with the graph state may be inferred from the stabilizer measurements.}
\label{fig:logrob_chain}
\end{figure}
\begin{figure}
\includegraphics[width=0.5\textwidth]{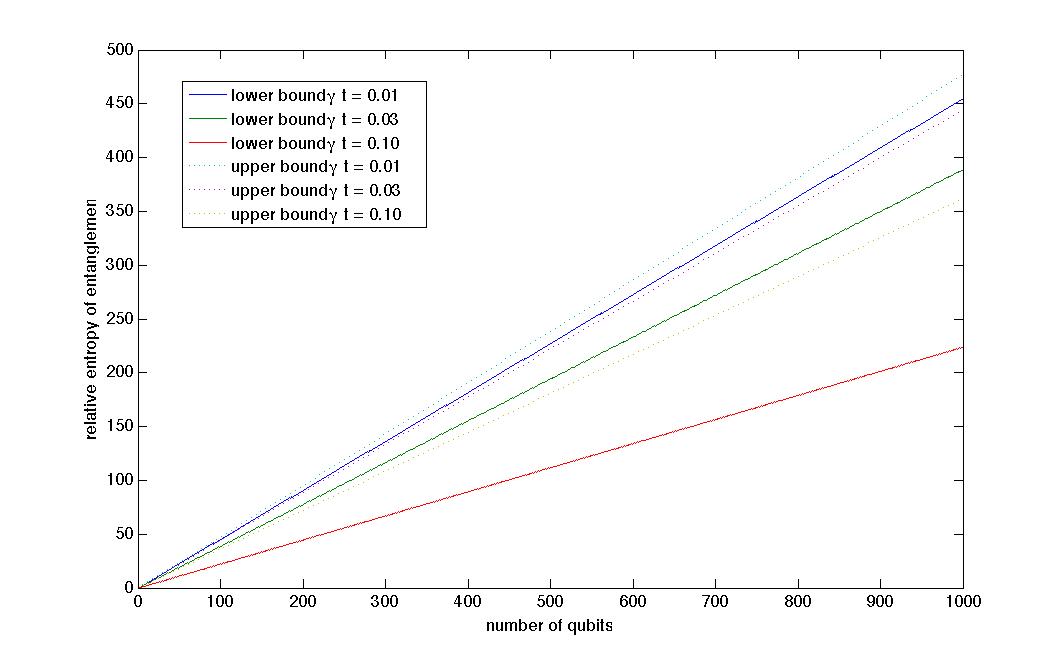}
\caption{Upper and lower bound on the estimate of the relative entropy of entanglement for linear graph states up to 1000 qubits subject to local dephasing. In contrast to the robustness of entanglement there is no limit to the lower bound, but the difference between upper and lower bounds grows with system size.}
\label{fig:relent_chain}
\end{figure}

\paragraph{Quality and scaling of the bounds} In order to check the quality of the entanglement estimates, we consider noisy two-colorable graph states. Assuming the experiments starts from a perfect graph state, which is then subjected to local dephasing for a certain time, we can take the density matrix time evolution to be governed by the following master equation:
\begin{equation}
\dot{\rho} = - \frac{\gamma}{2} (\sum_i Z_i \rho Z_i -\rho), 
\end{equation}
where $\gamma$ is the dephasing constant. The effect of such noise on graph states has been studied in detail in Ref. \cite{Hein05}. Due to the dephasing the stabilizer coefficients suffer a decay exponential in the dephasing constant. For our test we consider a linear chain of qubits subject to this noise. It can be shown that the stabilizer coefficients obey the following time evolution in this noise model: $c_{i_1 \dots i_n} (t)= exp(-\gamma t \sum_k i_k)$. Estimates according to the described methods for the logarithm of the global robustness of entanglement and the relative entropy of entanglement are shown in Figures \ref{fig:logrob_chain} and resp. \ref{fig:relent_chain}. The robustness of entanglement can be estimated up to a certain number of qubits, for which a non-zero fidelity can be inferred with the target state. This effectively sets a threshold to the estimation. In contrast, the relative entropy may be estimated even for larger noisy systems without suffering from the threshold problem. However, the difference between lower and upper bounds apparently grows with system size.

In many cases, stabilizer measurements are carried out via local measurements. This local information on the quantum state could in principle be used to improve the bounds on minimization of entanglement measures from incomplete information on the density matrix, since they restrict the set of separable states involved in the optimization. Note however, that local measurement operators generally do not commute with the stabilizer operators. This implies that symmetries cannot be exploited. For this reason we do not consider local measurement data in our scheme.
\paragraph{Conclusion}
Here we have shown how entanglement of arbitrarily large graph states can be estimated from simple measurements. High-quality bounds on the robustness of entanglement and the relative entropy of entanglement have been derived for stabilizer measurements. The stabilizers of two-colorable graph states can be measured in two measurement settings (assuming the measurements can be performed simultaneously), thus our scheme avoids the exponential overhead required by full-state tomography. In addition, the results presented here are of an analytical form that allows for extremely efficient post-processing. In contrast, quantum state tomography requires computationally hard post-processing of the measurement data to create an estimate of the real density matrix, and schemes for entanglement estimation from incomplete measurement data usually rely on numerical methods such as convex optimization which is limited to systems no larger than 20 qubits - even if symmetries can be exploited. Our scheme should therefore be invaluable for future graph-state experiments.

Our results may also be interesting to study the effects of various noise models on the entanglement dynamics of two-colorable graph states. One step in this direction has been made in Ref. \cite{Cavalcanti09}, where the entanglement of graph states under the influence of Pauli maps is investigated.

This work was supported by the EU Integrated Project QAP and EU STREP projects HIP and CORNER. MBP acknowledges an Alexander von Humboldt Professorship.


\begin{thebibliography}{99}
\expandafter\ifx\csname
natexlab\endcsname\relax\def\natexlab#1{#1}\fi
\expandafter\ifx\csname bibnamefont\endcsname\relax
  \def\bibnamefont#1{#1}\fi
\expandafter\ifx\csname bibfnamefont\endcsname\relax
  \def\bibfnamefont#1{#1}\fi
\expandafter\ifx\csname citenamefont\endcsname\relax
  \def\citenamefont#1{#1}\fi
\expandafter\ifx\csname url\endcsname\relax
  \def\url#1{\texttt{#1}}\fi
\expandafter\ifx\csname urlprefix\endcsname\relax\def\urlprefix{URL
}\fi \providecommand{\bibinfo}[2]{#2}
\providecommand{\eprint}[2][]{\url{#2}}

\bibitem[{\citenamefont{GŸhne and Toth}(2009)}]{GT09}
\bibinfo{author}{\bibfnamefont{O.} \bibnamefont{G\"uhne}} \bibnamefont{and}
\bibinfo{author}{\bibfnamefont{G.} \bibnamefont{T\'{o}th}} ,
  \bibinfo{journal}{Phys. Rep.} \textbf{\bibinfo{volume}{474}},
  \bibinfo{pages}{1} (\bibinfo{year}{2009}).

\bibitem[{\citenamefont{Plenio and Virmani}(2007)}]{PV07}
\bibinfo{author}{\bibfnamefont{M.~B.} \bibnamefont{Plenio}} \bibnamefont{and}
  \bibinfo{author}{\bibfnamefont{S.}~\bibnamefont{Virmani}},
  \bibinfo{journal}{Quant. Inf. Comp.} \textbf{\bibinfo{volume}{7}},
  \bibinfo{pages}{1} (\bibinfo{year}{2007}).

  \bibitem[{\citenamefont{Cramer and Plenio}(2010)}]{CP10}
\bibinfo{author}{\bibfnamefont{M.} \bibnamefont{Cramer}} \bibnamefont{and}
\bibinfo{author}{\bibfnamefont{M. B.} \bibnamefont{Plenio}} ,
  \bibinfo{e-print}{arXiv:1002.3780} 
 (\bibinfo{year}{2010}).
\bibinfo{author}{\bibfnamefont{S. T.} \bibnamefont{Flammia}},
\bibinfo{author}{\bibfnamefont{D.} \bibnamefont{Gross}},
\bibinfo{author}{\bibfnamefont{S. D.} \bibnamefont{Bartlett}}, and
\bibinfo{author}{\bibfnamefont{R.} \bibnamefont{Somma}}, 
 \bibinfo{e-print}{arXiv:1002.3839}  (\bibinfo{year}{2010}).
\bibinfo{author}{\bibfnamefont{O.} \bibnamefont{Landon-Cardinal}},
\bibinfo{author}{\bibfnamefont{Y.-K.} \bibnamefont{Liu}}, and
\bibinfo{author}{\bibfnamefont{D.} \bibnamefont{Poulin}},
 \bibinfo{e-print}{arXiv:1002.4632}  (\bibinfo{year}{2010}).

\bibitem[{\citenamefont{Greenberger et~al.}(1989)\citenamefont{Greenberger,
  Horne, and Zeilinger}}]{Greenberger89}
\bibinfo{author}{\bibfnamefont{D.~M.} \bibnamefont{Greenberger}},
  \bibinfo{author}{\bibfnamefont{M.~A.} \bibnamefont{Horne}}, \bibnamefont{and}
  \bibinfo{author}{\bibfnamefont{A.}~\bibnamefont{Zeilinger}},
  \emph{\bibinfo{title}{Bell's Theorem, Quantum Theory, and Conceptions of the
  Universe}} (\bibinfo{publisher}{M. Kafatos (Ed.) Kluwer Academic},
  \bibinfo{address}{Dordrecht}, \bibinfo{year}{1989}), pp.
  \bibinfo{pages}{69--72}.

\bibitem[{\citenamefont{Briegel and Raussendorf}(2001)}]{Briegel01}
\bibinfo{author}{\bibfnamefont{H.~J.} \bibnamefont{Briegel}} \bibnamefont{and}
  \bibinfo{author}{\bibfnamefont{R.}~\bibnamefont{Raussendorf}},
  \bibinfo{journal}{Phys. Rev. Lett.} \textbf{\bibinfo{volume}{86}},
  \bibinfo{pages}{910} (\bibinfo{year}{2001}).
  
\bibitem[{\citenamefont{Walther et al.}(2005)}]{Walther05}
\bibinfo{author}{\bibfnamefont{P.} \bibnamefont{Walther}} \bibnamefont{et al.}
  \bibinfo{journal}{Nature} \textbf{\bibinfo{volume}{434}},
  \bibinfo{pages}{169} (\bibinfo{year}{2005}).
%
\bibitem[{\citenamefont{Kiesel et al.}(2005)}]{Kiesel05}
\bibinfo{author}{\bibfnamefont{N.} \bibnamefont{Kiesel}} \bibnamefont{et al.}
  \bibinfo{journal}{Phys. Rev. Lett.} \textbf{\bibinfo{volume}{95}},
  \bibinfo{pages}{210502} (\bibinfo{year}{2005}).
%
\bibitem[{\citenamefont{Lu et al.}(2007)}]{Lu07}
\bibinfo{author}{\bibfnamefont{C. Y.} \bibnamefont{Lu.}} \bibnamefont{et al.}
  \bibinfo{journal}{Nature Phys.} \textbf{\bibinfo{volume}{3}},
  \bibinfo{pages}{91} (\bibinfo{year}{2007}).
%
\bibitem[{\citenamefont{Chen et al.}(2007)}]{Chen07}
\bibinfo{author}{\bibfnamefont{K.} \bibnamefont{Chen}} \bibnamefont{et al.}
  \bibinfo{journal}{Phys. Rev. Lett.} \textbf{\bibinfo{volume}{99}},
  \bibinfo{pages}{120503} (\bibinfo{year}{2007}).
%
\bibitem[{\citenamefont{Vallone et al.}(2008)}]{Vallone08}
\bibinfo{author}{\bibfnamefont{G.} \bibnamefont{Vallone}},
\bibinfo{author}{\bibfnamefont{E.} \bibnamefont{Pomarico}},
\bibinfo{author}{\bibfnamefont{F.} \bibnamefont{De Martini}}, \bibnamefont{and}
\bibinfo{author}{\bibfnamefont{P.} \bibnamefont{Mataloni}},
  \bibinfo{journal}{Phys. Rev. Lett.} \textbf{\bibinfo{volume}{99}},
  \bibinfo{pages}{120503} (\bibinfo{year}{2007}).

\bibitem[{\citenamefont{Mandel et al.}(2003)}]{Mandel03}
\bibinfo{author}{\bibfnamefont{O.} \bibnamefont{Mandel}},
\bibinfo{author}{\bibfnamefont{M.} \bibnamefont{Greiner}}, 
\bibinfo{author}{\bibfnamefont{A.} \bibnamefont{Widera}}, 
\bibinfo{author}{\bibfnamefont{T.} \bibnamefont{Rom}}, 
\bibinfo{author}{\bibfnamefont{T. W.} \bibnamefont{H\"ansch}} 
\bibnamefont{and}
\bibinfo{author}{\bibfnamefont{I.} \bibnamefont{Bloch}},
  \bibinfo{journal}{Nature (London)} \textbf{\bibinfo{volume}{425}},
  \bibinfo{pages}{937} (\bibinfo{year}{2003}).

\bibitem[{\citenamefont{Wunderlich, Wunderlich, Singer, Schmidt-Kaler}(2009)}]{WWSSK09}
\bibinfo{author}{\bibfnamefont{H.} \bibnamefont{Wunderlich}},
\bibinfo{author}{\bibfnamefont{Chr.} \bibnamefont{Wunderlich}}, 
\bibinfo{author}{\bibfnamefont{K.} \bibnamefont{Singer}} 
\bibnamefont{and}
\bibinfo{author}{\bibfnamefont{F.} \bibnamefont{Schmidt-Kaler}},
  \bibinfo{journal}{Phys. Rev. A} \textbf{\bibinfo{volume}{79}},
  \bibinfo{pages}{052324} (\bibinfo{year}{2009}).
%
\bibitem[{\citenamefont{Stock and James}(2009)}]{Stock09}
\bibinfo{author}{\bibfnamefont{R.} \bibnamefont{Stock}} \bibnamefont{and}
\bibinfo{author}{\bibfnamefont{D. F.} \bibnamefont{James}} ,
  \bibinfo{journal}{Phys. Rev. Lett.} \textbf{\bibinfo{volume}{102}},
  \bibinfo{pages}{170501} (\bibinfo{year}{2009}).
%
\bibitem[{\citenamefont{Ivanov et al.}(2008)}]{Ivanov08}
\bibinfo{author}{\bibfnamefont{P. A.} \bibnamefont{Ivanov}},
\bibinfo{author}{\bibfnamefont{N. V.} \bibnamefont{Vitanov}} \bibnamefont{and}
\bibinfo{author}{\bibfnamefont{M. B.} \bibnamefont{Plenio}},
  \bibinfo{journal}{Phys. Rev. A} \textbf{\bibinfo{volume}{78}},
  \bibinfo{pages}{12323} (\bibinfo{year}{2008}).


\bibitem[{\citenamefont{Audenaert and Plenio}(2006)}]{AP06}
\bibinfo{author}{\bibfnamefont{K.~M.~R.} \bibnamefont{Audenaert}} \bibnamefont{and}
  \bibinfo{author}{\bibfnamefont{M.~B.}~\bibnamefont{Plenio}},
  \bibinfo{journal}{New J. Phys.} \textbf{\bibinfo{volume}{8}},
  \bibinfo{pages}{266} (\bibinfo{year}{2006}).
\bibinfo{author}{\bibfnamefont{J.} \bibnamefont{Eisert}},
\bibinfo{author}{\bibfnamefont{F.~G.~S.~L.} \bibnamefont{Brand\~{a}o}}, \bibnamefont{and}
  \bibinfo{author}{\bibfnamefont{K.~M.~R.}~\bibnamefont{Audenaert}},
  \bibinfo{journal}{New J. Phys.} \textbf{\bibinfo{volume}{9}},
  \bibinfo{pages}{46} (\bibinfo{year}{2007}).   
\bibinfo{author}{\bibfnamefont{O.} \bibnamefont{G\"uhne}},
\bibinfo{author}{\bibfnamefont{M.} \bibnamefont{Reimpell}}, \bibnamefont{and}
  \bibinfo{author}{\bibfnamefont{R. F.} \bibnamefont{Werner}},
  \bibinfo{journal}{Phys. Rev. Lett.} \textbf{\bibinfo{volume}{98}},
  \bibinfo{pages}{110502} (\bibinfo{year}{2007}).

\bibitem[{\citenamefont{Wunderlich and Plenio}(2009)}]{WP0902}
\bibinfo{author}{\bibfnamefont{H.} \bibnamefont{Wunderlich}} \bibnamefont{and}
  \bibinfo{author}{\bibfnamefont{M.~B.}~\bibnamefont{Plenio}},
  \bibinfo{journal}{J. Mod. Opt.} \textbf{\bibinfo{volume}{56}},
  \bibinfo{pages}{2100} (\bibinfo{year}{2009}).

\bibitem[{\citenamefont{Vedral and Plenio}(1998)}]{VP98}
\bibinfo{author}{\bibfnamefont{V.} \bibnamefont{Vedral}} \bibnamefont{and}
  \bibinfo{author}{\bibfnamefont{M.~B.}~\bibnamefont{Plenio}},
  \bibinfo{journal}{Phys. Rev. A} \textbf{\bibinfo{volume}{57}},
  \bibinfo{pages}{1619} (\bibinfo{year}{1998}).
  
\bibitem[{\citenamefont{Vidal, Tarrach, Harrow, Nielsen, Steiner}(1999)}]{RobustnessEnt}
\bibinfo{author}{\bibfnamefont{G.} \bibnamefont{Vidal}} \bibnamefont{and}
  \bibinfo{author}{\bibfnamefont{R.}~\bibnamefont{Tarrach}},
  \bibinfo{journal}{Phys. Rev. A} \textbf{\bibinfo{volume}{59}},
  \bibinfo{pages}{141} (\bibinfo{year}{1999}).
\bibinfo{author}{\bibfnamefont{A.} \bibnamefont{Harrow}} \bibnamefont{and}
  \bibinfo{author}{\bibfnamefont{M.~A.}~\bibnamefont{Nielsen}},
  \bibinfo{journal}{Phys. Rev. A} \textbf{\bibinfo{volume}{68}},
  \bibinfo{pages}{012308} (\bibinfo{year}{2003}).
\bibinfo{author}{\bibfnamefont{M.} \bibnamefont{Steiner}},
  \bibinfo{journal}{Phys. Rev. A} \textbf{\bibinfo{volume}{67}},
  \bibinfo{pages}{054305} (\bibinfo{year}{2003}).

\bibitem[{\citenamefont{Shimony, Barnum, Linden, Miyake, Wadati, Wei, Goldbart}(1995)}]{GeoEnt}
\bibinfo{author}{\bibfnamefont{A.} \bibnamefont{Shimony}},
  \bibinfo{journal}{Ann. NY Acad. Sci.} \textbf{\bibinfo{volume}{755}},
  \bibinfo{pages}{675} (\bibinfo{year}{1995}).
\bibinfo{author}{\bibfnamefont{H.} \bibnamefont{Barnum}} \bibnamefont{and}
  \bibinfo{author}{\bibfnamefont{N.}~\bibnamefont{Linden}},
  \bibinfo{journal}{J. Phys. A} \textbf{\bibinfo{volume}{34}},
  \bibinfo{pages}{6787} (\bibinfo{year}{2001}).
\bibinfo{author}{\bibfnamefont{T.-C.} \bibnamefont{Wei}} \bibnamefont{and}
  \bibinfo{author}{\bibfnamefont{P.~M.}~\bibnamefont{Goldbart}},
  \bibinfo{journal}{Phys. Rev. A} \textbf{\bibinfo{volume}{68}},
  \bibinfo{pages}{042307} (\bibinfo{year}{2003}).

\bibitem[{\citenamefont{Markham, Miyake, and Virmani}(2007)}]{MMV07}
\bibinfo{author}{\bibfnamefont{D.} \bibnamefont{Markham}}, 
  \bibinfo{author}{\bibfnamefont{A.} \bibnamefont{Miyake}}, \bibnamefont{and}
	\bibinfo{author}{\bibfnamefont{S.} \bibnamefont{Virmani}},
  \bibinfo{journal}{New J. Phys.} \textbf{\bibinfo{volume}{9}},
  \bibinfo{pages}{194} (\bibinfo{year}{2007}).
%
\bibitem[{\citenamefont{Wunderlich and Plenio}(2009)}]{WP0907}
\bibinfo{author}{\bibfnamefont{H.} \bibnamefont{Wunderlich}} \bibnamefont{and}
  \bibinfo{author}{\bibfnamefont{M.~B.}~\bibnamefont{Plenio}},
  \bibinfo{journal}{Int. J. Quant. Inf.} (in Press)
 \bibinfo{e-print}{arxiv:quant-ph/0907.1848} 
(\bibinfo{year}{2009}).
%
\bibitem[{\citenamefont{Hein, Duer, Briegel}(2005)}]{Hein05}
\bibinfo{author}{\bibfnamefont{M.} \bibnamefont{Hein}},
\bibinfo{author}{\bibfnamefont{W.} \bibnamefont{D\"ur}},
 \bibnamefont{and}
  \bibinfo{author}{\bibfnamefont{H.~J.}~\bibnamefont{Briegel}},
  \bibinfo{journal}{Phys. Rev. A} \textbf{\bibinfo{volume}{71}},
  \bibinfo{pages}{032350} (\bibinfo{year}{2005}).
%
\bibitem[{\citenamefont{Plenio, Virmani, Papadopolous}(2000)}]{PVP00}
\bibinfo{author}{\bibfnamefont{M. B.} \bibnamefont{Plenio}},
\bibinfo{author}{\bibfnamefont{S.} \bibnamefont{Virmani}},
 \bibnamefont{and}
  \bibinfo{author}{\bibfnamefont{P.}~\bibnamefont{Papadopolous}},
  \bibinfo{journal}{J. Phys. A} \textbf{\bibinfo{volume}{33}},
  \bibinfo{pages}{L193-197} (\bibinfo{year}{2000}).
%
\bibitem[{\citenamefont{Cavalcanti et al.}(2009)}]{Cavalcanti09}
\bibinfo{author}{\bibfnamefont{D.} \bibnamefont{Cavalcanti}},
\bibinfo{author}{\bibfnamefont{R.} \bibnamefont{Chaves}},
\bibinfo{author}{\bibfnamefont{L.} \bibnamefont{Aolita}},
\bibinfo{author}{\bibfnamefont{A.} \bibnamefont{Davidovich}},
 \bibnamefont{and}
  \bibinfo{author}{\bibfnamefont{A.}~\bibnamefont{Acin}},
  \bibinfo{journal}{Phys. Rev. Lett.} \textbf{\bibinfo{volume}{103}},
  \bibinfo{pages}{030502} (\bibinfo{year}{2009}).
%
\end{thebibliography}
\end{document}